%% file: main.tex
\documentclass[10pt,twocolumn,letterpaper]{article}

\usepackage[pagenumbers]{cvpr} 

\usepackage{graphicx}
\usepackage{amsmath}
\usepackage{amssymb}
\usepackage{booktabs}

\usepackage[pagebackref,breaklinks,colorlinks]{hyperref}

\usepackage[capitalize]{cleveref}
\crefname{section}{Sec.}{Secs.}
\Crefname{section}{Section}{Sections}
\Crefname{table}{Table}{Tables}
\crefname{table}{Tab.}{Tabs.}

\def\cvprPaperID{29} 
\def\confName{CVPR}
\def\confYear{2022}

\newcommand{\comment}[1]{}

\begin{document}

\title{Searching for Efficient Neural Architectures for On-Device ML on Edge TPUs}

\author{
Berkin Akin\\
{\tt\small bakin@google.com}
\and
Suyog Gupta\\
{\tt\small suyoggupta@google.com}
\and
Yun Long\\
{\tt\small longy@google.com}
\and
Anton Spiridonov\\
{\tt\small tohaspiridonov@google.com}
\and
Zhuo Wang\\
{\tt\small zhuowang@google.com}
\and
Marie White\\
{\tt\small mariewhite@google.com}
\and
Hao Xu\\
{\tt\small iamxuhao@google.com}
\and
Ping Zhou\\
{\tt\small zhouping@google.com}
\and
Yanqi Zhou\\
{\tt\small yanqiz@google.com}
}

\maketitle

\input{0_abstract}
\input{1_intro}
\input{2_related_work}

\input{3__search_infra}
\input{3_search_space}
\input{4_models}

\input{6_conclusion}

{\small
\bibliographystyle{ieee_fullname}
\bibliography{egbib}
}

\end{document}

%% file: 0_abstract.tex
\begin{abstract}
On-device ML accelerators are becoming a standard in modern mobile system-on-chips (SoC).
Neural architecture search (NAS) comes to the rescue for efficiently utilizing the high compute throughput offered by these accelerators. However, existing NAS frameworks have several practical limitations 
in scaling to multiple tasks and different target platforms.
In this work, we provide a two-pronged approach to this challenge:
(i) a NAS-enabling infrastructure that decouples model cost evaluation, search space design, and the NAS algorithm to rapidly target various on-device ML tasks, and
(ii) search spaces crafted from group convolution based inverted bottleneck (IBN) variants that provide flexible quality/performance trade-offs on ML accelerators, 
complementing the existing full and depthwise convolution based IBNs.
Using this approach we target a state-of-the-art mobile platform, Google Tensor SoC, 
and demonstrate neural architectures\footnote{\url{https://github.com/tensorflow/models/tree/master/official/projects/edgetpu/}} that improve the quality-performance pareto frontier for various computer vision (classification, detection, segmentation) as well as natural language processing tasks.
\end{abstract}

%% file: 1_intro.tex
\section{Introduction}
\label{sec:intro}
Due to the diminishing returns in the performance gains with the technology scaling in the post-Moore era, specialized ML accelerators became an essential component in most of the modern mobile system-on-chip (SoC) platforms to serve the needs of the real-time on-device ML workloads. 
Specialized ML accelerators (such as TPUs \cite{tensor_soc}, NPUs \cite{apple_ne, samsung_npu}) provide a substantial peak computation throughput, however the neural networks can extract optimal performance
only when they are co-designed for the underlying hardware architecture.

There has been significant effort in hand crafting optimized neural architectures for specific target platforms  \cite{mobilenet_cvpr_howard, mobilenet2_cvpr_sandler}. However, the increased complexity of the neural models and the variety of the target platforms gave rise to automated neural architecture search (NAS/AutoML) approaches \cite{mnas_fpn_cvpr_chen, tan2019mnasnet, cai2019proxylessnas, one_shot, tunas, liu2019darts}.
Although there are a variety of NAS frameworks, there are practical scalability limitations when it comes to
designing neural architectures for different task domains and/or target platforms.

\begin{figure}
    \centering
    \includegraphics[width=3.33in]{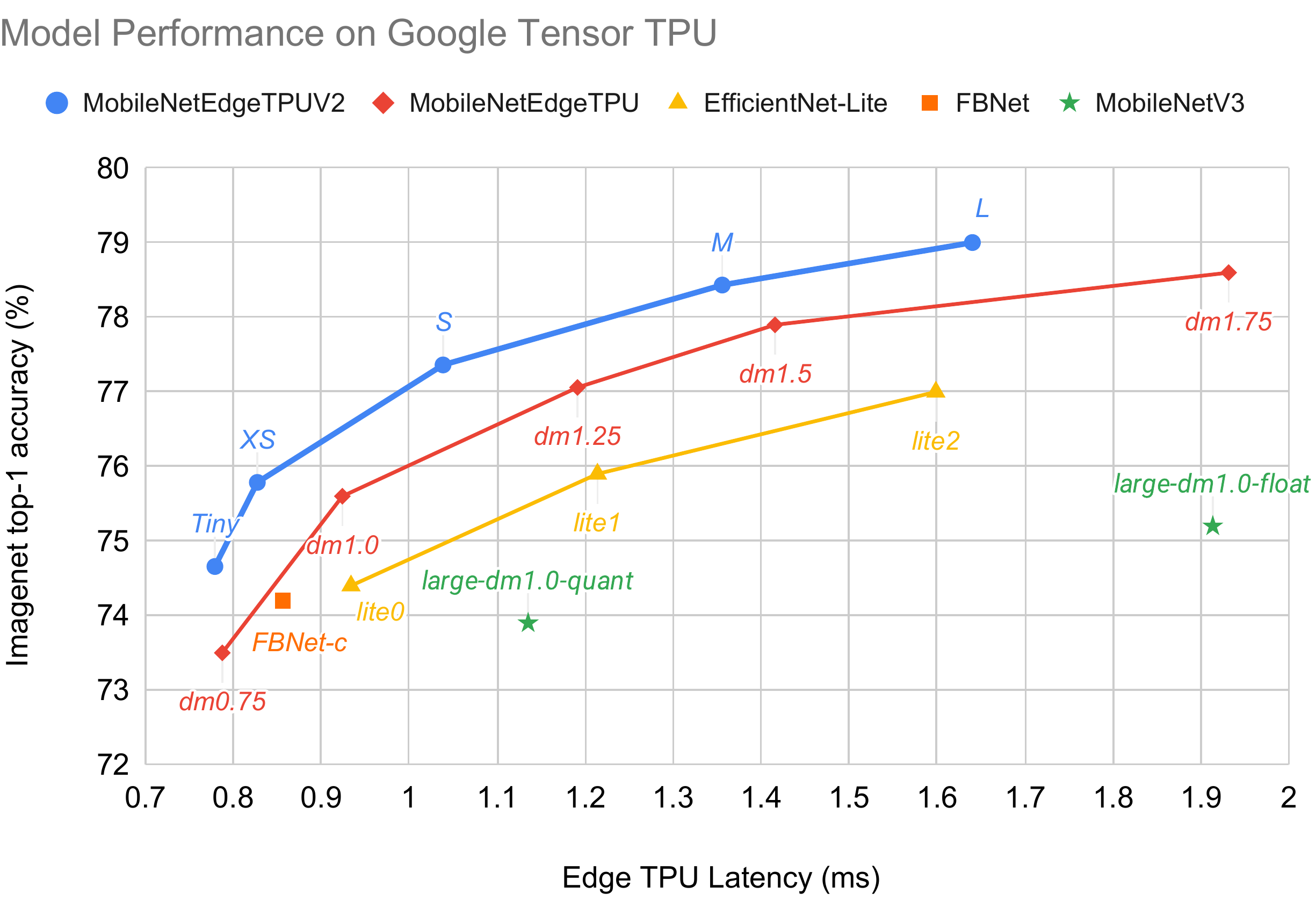}
    \caption{Proposed MobileNetEdgeTPUV2 models achieve higher ImageNet top-1 accuracy at lower latency when running on Google Tensor’s TPU \cite{tensor_soc} compared to MobileNetEdgeTPU, EfficientNet, FBNet, MobilenetV3. All models are quantized unless noted otherwise.}
    \label{fig:classify_tpu}
\end{figure}

Either using NAS \cite{mobiledet_cvpr_xiong, mobilenet_edge_paper, wu2019fbnet} or through manual design  \cite{mobilenet2_cvpr_sandler, resnext}, inverted bottleneck (IBN) layers have been predominant in building computer vision models. Although conventional IBNs that use depthwise convolutions have been very successful for mobile CPUs, prior work highlighted the use of full convolutions can significantly improve the model's accuracy-latency trade-off \cite{mobiledet_cvpr_xiong, mobilenet_edge_paper}. 
Moreover, using full convolutions in IBNs allow \emph{fusing} the pointwise expansion with the main full convolutions that can enable further latency optimizations on ML accelerators such as Edge TPUs \cite{mobilenet_edge_blog, mobilenet_edge_paper}.
However, fused-IBNs can have substantially high computational and memory requirements for spatially narrow, channel-wise deep tensor shapes that are typical in the later stages of vision models, limiting their use throughout the model and leaving the depthwise-IBN as the only alternative.

We make an observation that a key factor in the high hardware efficiency of full convolutions on ML accelerators is the increased data reuse due to channel-wise convolutions.
Depthwise separable convolutions remove the channel-wise convolution dimension which reduces the overall parameter and operation count but at the same time leading to extremely low hardware utilizations.
We propose group convolution (GC) based IBNs, where the channel-wise convolution is still performed but limited to within each group. 
This allows GC-IBN variants to reach hardware utilization levels similar to 
the full convolution based IBNs but with much fewer parameter/operation counts.

Moreover, to address the practical limitations of NAS frameworks, we built a scalable infrastructure which decouples the search space design, platform cost evaluation (e.g. latency, energy) and the NAS algorithm. We provide cost evaluation as a gRPC\cite{grpc} service where multi-trial or one-shot NAS clients can plug into, either for directly evaluating a search candidate (multi-trial) or to build learned cost models (one-shot). 

As a concrete case study, we use the proposed infrastructure on the search spaces including the proposed GC-based IBNs and target the Edge TPU ML accelerator in the Google Tensor mobile SoC \cite{tensor_soc} for the on-device ML tasks identified by MLPerf Mobile Inference suite \cite{mlperf_mobile} which includes image classification, object detection, semantic segmentation and natural language processing. For each task, we demonstrate that the models designed by using our framework significantly improves the accuracy-performance pareto-frontier through the latency and energy measurements from Pixel 6 devices.

%% file: 2_related_work.tex
\section{Related Work}
\label{sec:related}

{\bf Neural Architecture Search (NAS)} was proposed to automate the design of neural network architectures, 
often aiming to improve model quality given a cost metric \cite{mnas_fpn_cvpr_chen, tan2019mnasnet, cai2019proxylessnas, one_shot, tunas, liu2019darts}. 
Since evaluating a candidate model's quality requires expensive training jobs in a \emph{multi-trial} NAS \cite{tan2019mnasnet}, one-shot approaches with weight sharing in a super-network are proposed \cite{cai2019proxylessnas, liu2019darts, tunas}. 
In this work we are not proposing a new NAS algorithm. 
Rather we are making an observation that various NAS methods and their implementations come 
with either algorithmic or practical limitations/benefits. 
We build an infrastructure which can interface with various NAS clients and 
exercise it for rapid development targeting a state-of-the-art platform 
for multiple on-device ML tasks from different domains.

{\bf Inverted bottleneck} blocks (IBN) have been used extensively in building computer vision models \cite{mobilenet2_cvpr_sandler, resnext, howard2017mobilenets, mobilenet_cvpr_howard, mobiledet_cvpr_xiong, wu2019fbnet}. 
Conventionally the use of depthwise separable convolutions along with separate pointwise expansion and projection has shown to be very effective for mobile CPUs \cite{howard2017mobilenets, mobilenet_cvpr_howard, mobilenet2_cvpr_sandler}. Recent work also showed that using full convolutions where expansion and the $K\times K$ kernel is fused can be very efficient on ML accelerators \cite{mobiledet_cvpr_xiong, mobilenet_edge_blog, mobilenet_edge_paper}.

{\bf Group convolutions} (GC) were originally intended for model parallelism across GPUs in AlexNet \cite{alexnet}, yet they were also used as part of the IBN blocks \cite{zhang2017shufflenet, resnext, wu2019fbnet} to improve model quality. Recent FBNets \cite{wu2019fbnet} use GC in pointwise convolutions while keeping depthwise convolutions. ResNext \cite{resnext} divides the ResNet bottleneck blocks into groups, while ShuffleNet also uses shuffle operations to add cross-group feature exchange. In this work, we propose flexible GC based IBN variants that use GC as the $K\times K$ kernel and optionally keep the pointwise full convolutions.
We exploit the flexibility of GC to implement the expansion/projection as part of the $K\times K$ GC kernel and achieve fused GC IBNs similar to fused full convolution versions. 
We demonstrate that using GC IBNs opens up the search space between depthwise and full convolution based IBNs,
and create unique opportunities for efficient execution on ML accelerators.

%% file: 3__search_infra.tex
\section{Neural Architecture Search Infrastructure}
\label{sec:search_infra}
In this section, we introduce a scalable infrastructure we built to perform 
neural architecture search for optimizing various models on a dedicated ML accelerator (Edge TPU).
There are two major challenges to address with this infrastructure. 
First, different from optimizing models for CPUs, 
performance and power metrics of a model are harder to predict 
directly from the number of operations/parameters on ML accelerators.
With the software managed memory hierarchies of ML accelerators, 
achieved performance highly depend on how compiler maps the neural networks on the hardware.
Therefore, we need a way to collect accurate performance and power evaluations (PPE) for guiding the search.
Second, since we target diverse applications, 
the framework should unify the search space description and exploration 
flow and scale to different model domains.

\begin{figure}
    \centering
    \includegraphics[width=3.3in]{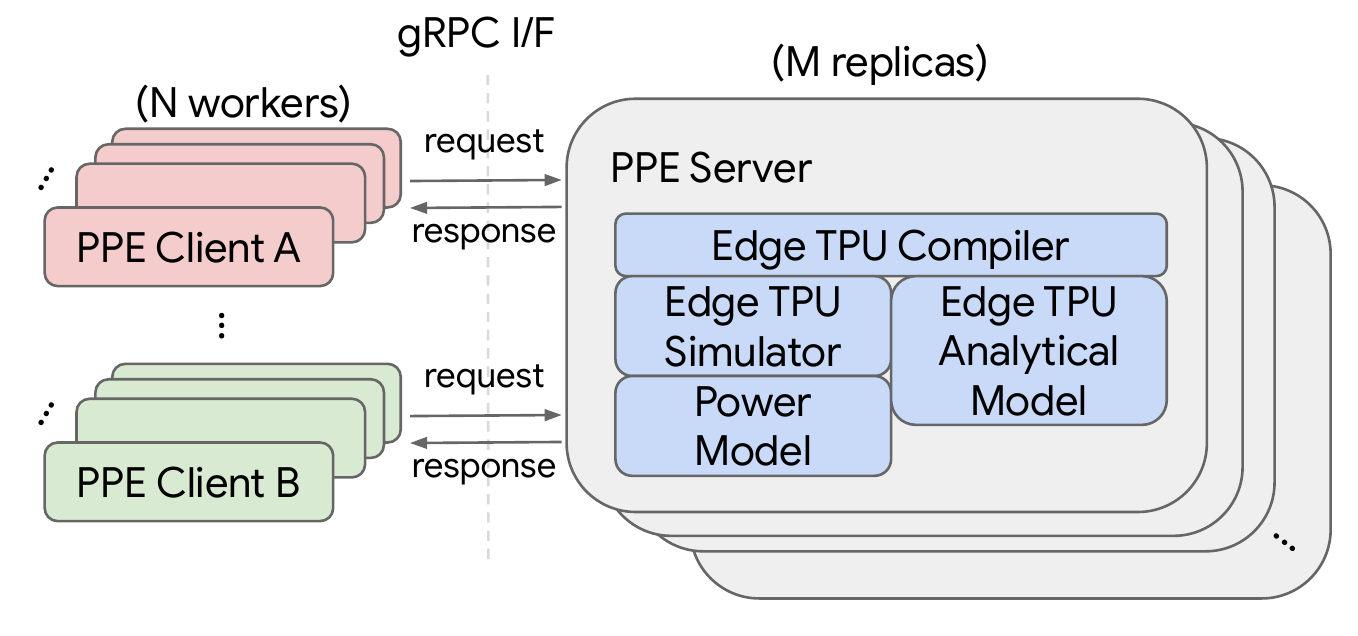}
    \caption{PPE service for model power/performance evaluation. 
    Multiple clients can have many workers serviced by several server replicas.}
    \label{fig:ppa-service}
\end{figure}

\subsection{Performance Power Evaluation (PPE) Service}
Figure~\ref{fig:ppa-service} shows the components of the PPE service. The server integrates the Edge TPU compiler, a cycle-accurate simulator, an analytical performance model for fast yet less accurate model simulation and a power estimator.
Clients can send independent estimation requests to the server via gRPC \cite{grpc} interface for a candidate model. 
The server is scaled to thousands of machines, so several requests can be served at the same time 
to serve the needs for highly parallel NAS.

\begin{figure}
    \centering
\includegraphics[width=3.3in]{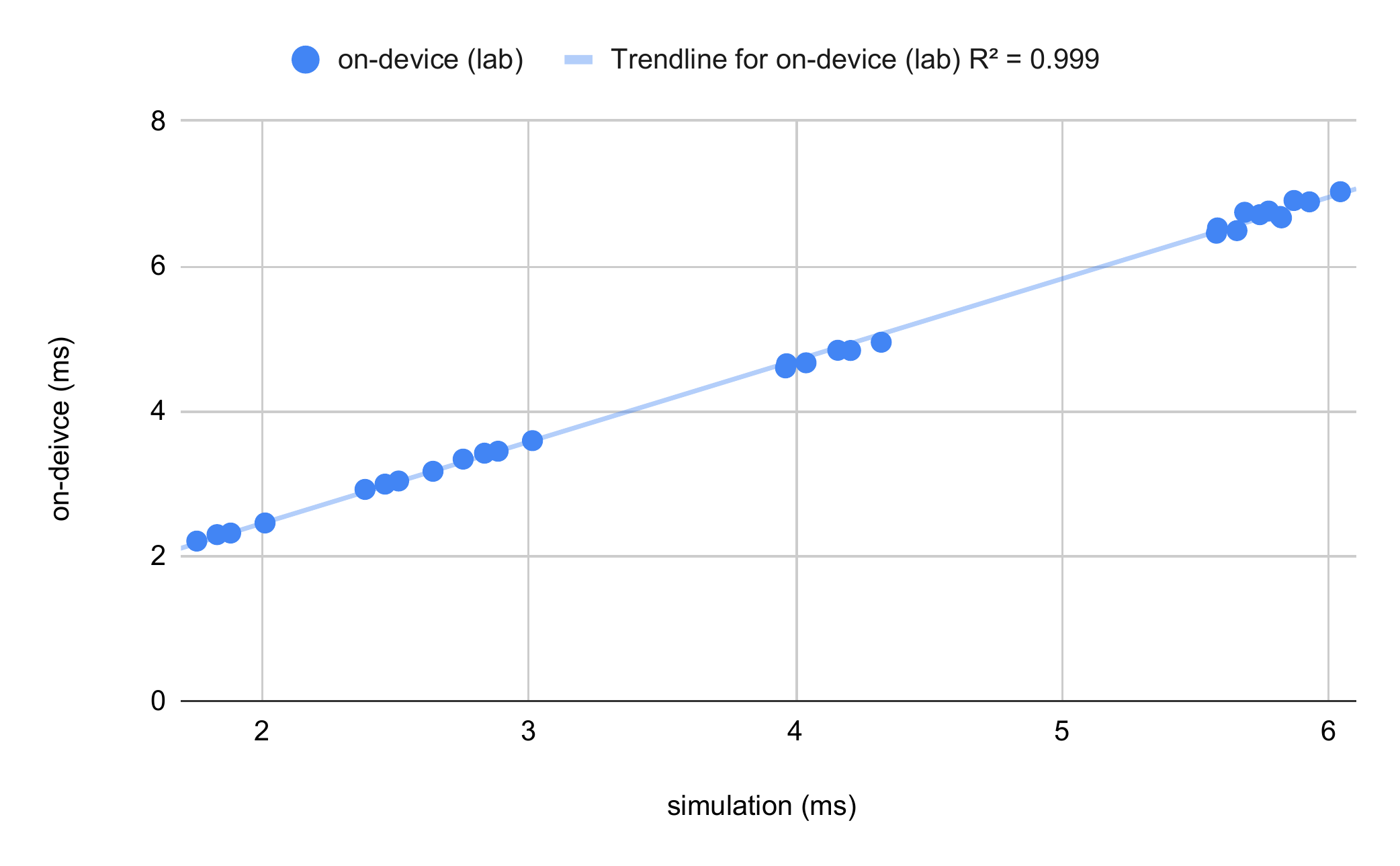}
    \caption{Comparing on-device latency measurement with the PPE service results.}
    \label{fig:estimation-validity}
\end{figure}

Figure~\ref{fig:estimation-validity} presents the correlation of on-device latency versus the latency from PPE service for randomly selected real use-case models. 
We observe that PPE latency is in general lower compared to the on-device latency, due to simulator's optimistic assumptions on system resources such as DRAM bandwidth. 
However, there is a very strong linear correlation between the PPE results and the real-device measurements ($R^2=0.99$).
This leads to correct relative ranking of the models which is the most critical in NAS.

\begin{figure}
    \centering
\includegraphics[width=3.3in]{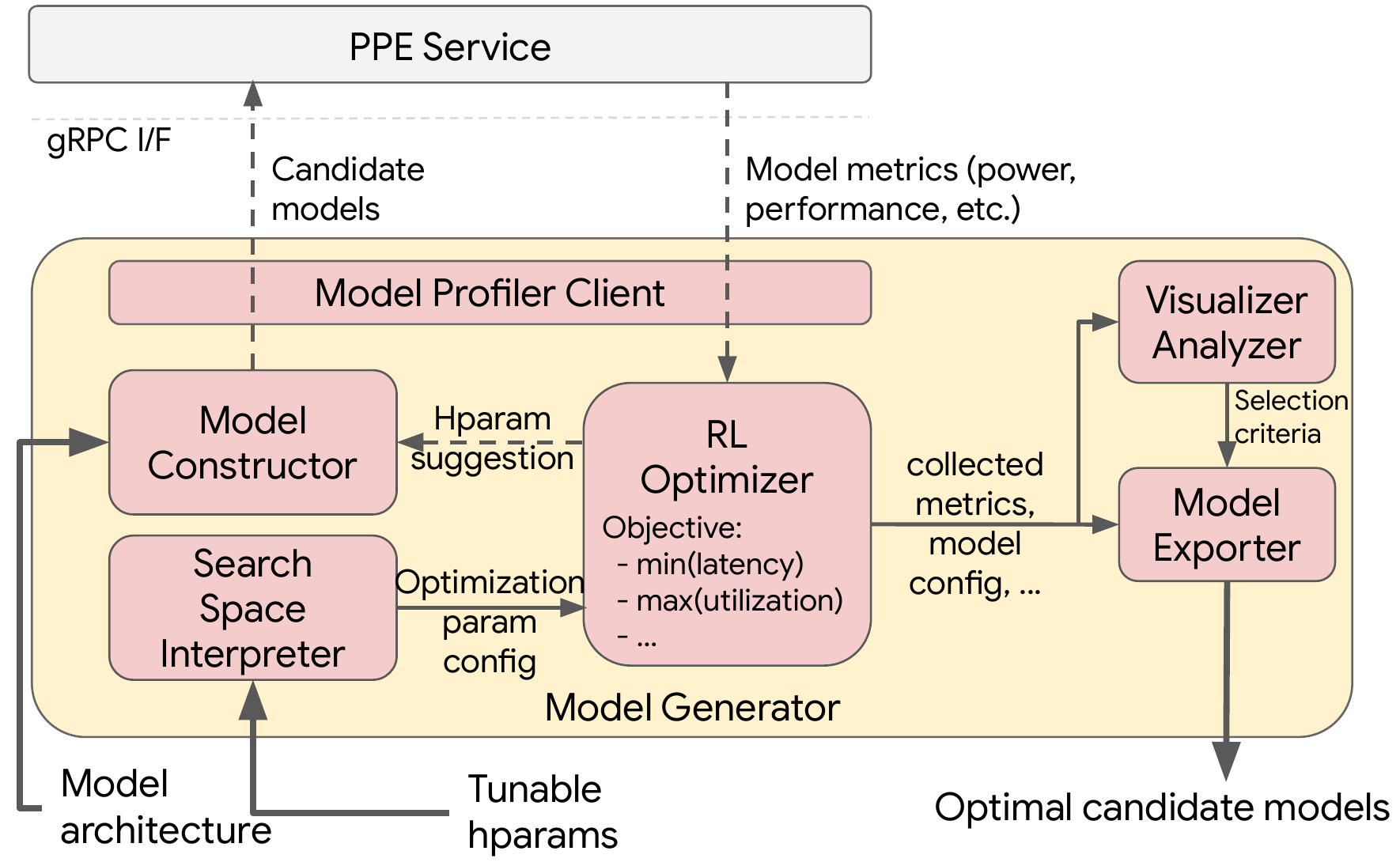}
    \caption{Model generator framework.}
    \label{fig:model-generator}
\end{figure}

\subsection{NAS Integration}
PPE service can be integrated with different NAS backends. In this work, we have utilized both one-shot and multi-trial NAS algorithms. For one-shot NAS, we used a weight sharing method based on TuNAS framework \cite{tunas}.
We have leveraged the one-shot NAS for classification and detection tasks. 
However, mostly due to practical challenges of searching for end-to-end models (backbone and head) and
compatibility of the neural operations (e.g. in transformer blocks) with weight sharing, 
we used multi-trial NAS for segmentation and NLP tasks.
However, note that this is not a fundamental limitation but rather an implementation decision 
for rapid deployment targeting a state-of-the-art platform.

{\bf Model Generator.} For the multi-trial approach, we have designed a model generator framework which is platform and task domain agnostic. Figure~\ref{fig:model-generator} is an overview of the framework. 
A user-friendly interface allows defining flexible search spaces by specifying model architecture topology and searchable parameters.
A reinforcement learning (RL) back-end takes in searchable parameters and provides trial suggestions in iterations.
The suggested candidate models are constructed and fed to PPE server as estimation requests.
The PPE server responses with the model metrics are used to train the Vizier-based \cite{vizier} RL agent
for optimizing the back-end to make the next iteration of suggestions towards an optimization goal (latency, power, model size, etc.).
A visualization and analysis tool is also integrated for assisting 
the selection of the best candidate models from the pool.
The user can then export the selected models in preferred formats for further evaluation (e.g. training) and deployment.
We use model generator as (i) a multi-trial NAS agent and (ii) an inverted bottleneck based neural block analyzer (see Section~\ref{sec:search_space}).

%% file: 3_search_space.tex
\section{Neural Architecture Search Space}
\label{sec:search_space}

\subsection{Inverted Bottlenecks (IBN)}
\begin{figure}
    \centering
    \includegraphics[width=3.3in]{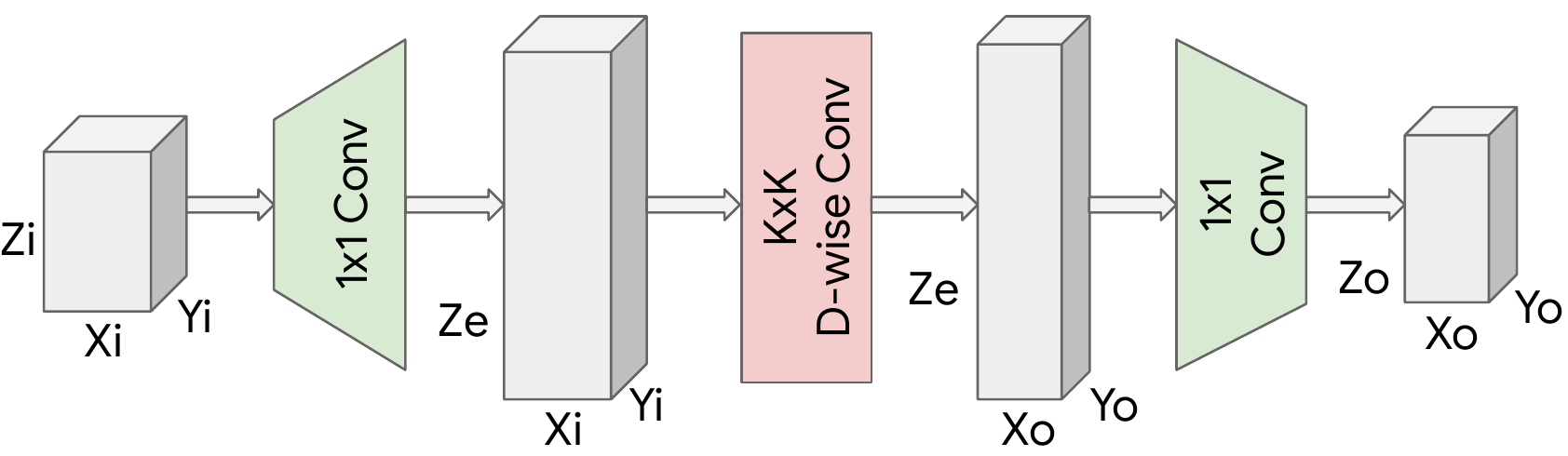}
    \caption{IBN using depthwise convolution \cite{mobilenet2_cvpr_sandler} (Depthwise-IBN).}
    \label{fig:ibn_dw}
\end{figure}

Inverted bottleneck layers, commonly abbreviated as IBNs, have been a predominant building
block in state-of-the-art computer vision models for mobile platforms \cite{mobiledet_cvpr_xiong, mobilenet_cvpr_howard, mobilenet2_cvpr_sandler}. The concept of a (inverted) bottleneck have also been extended to design of edge-device friendly NLP models \cite{sun2020mobilebert}. 
As shown in Figure~\ref{fig:ibn_dw}, a conventional IBN features a point-wise ($1\times 1$) convolution
that expands the input channel dimension to a larger value before applying a $K\times K$ depthwise convolution on the spatial dimensions. Finally, another point-wise convolution is used to project the expanded channel dimension to the desired final value. 

IBNs are originally designed for mobile CPUs to reduce the overall operation count in FLOPS (floating-point operations), the number of trainable parameters and improve hardware efficiency. The separation of convolutions along the channel and spatial dimension serves this goal compared to performing a full convolution at the expanded channel dimension. However, also observed by prior work \cite{mobiledet_cvpr_xiong, mobilenet_edge_paper}, not all FLOPS have the same efficiency, especially on mobile ML accelerators, where a regular convolution may run $3\times$ as fast on Edge TPUs than a
depthwise convolution even with $7\times$ as many FLOPS.

\begin{figure}
    \centering
\includegraphics[width=2.5in]{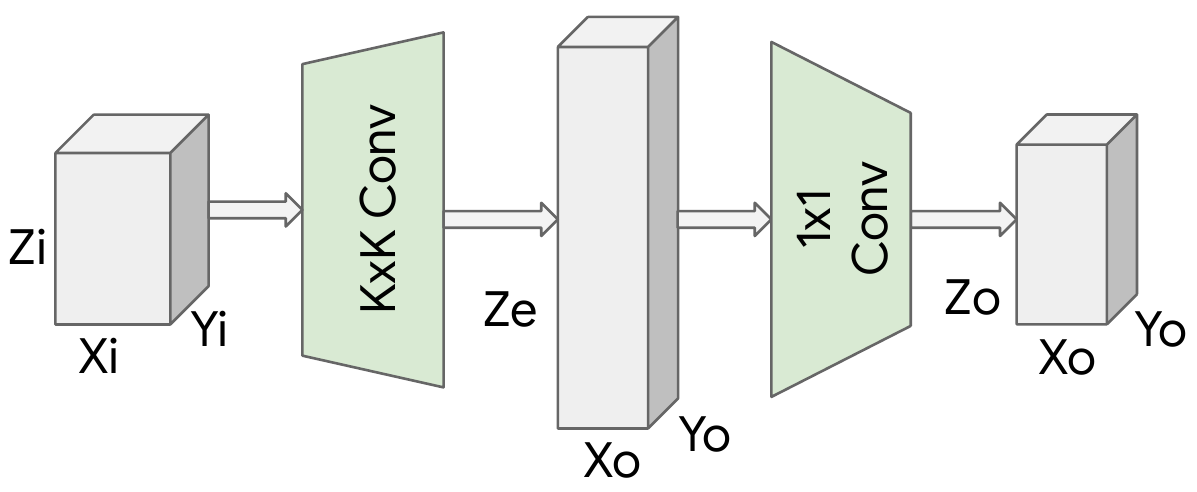}
    \caption{IBN using full convolution for the fused expansion and main kernel \cite{mobiledet_cvpr_xiong, mobilenet_edge_paper} (Fused-IBN).}
    \label{fig:ibn_fused}
\end{figure}

Motivated by this observation, Fused-IBN variants as shown in Figure~\ref{fig:ibn_fused} uses a regular full convolution instead of a separate pointwise expansion and a depthwise convolution kernel. Neural architecture search spaces augmented with the Fused-IBN were shown to improve model quality/latency trade-off for object detection \cite{mobiledet_cvpr_xiong} and image classification \cite{mobilenet_edge_paper} tasks.

Although the Fused-IBN variant can provide an efficient alternative to Depthwise-IBN, we observed that Fused-IBN were primarily used in the early layers of the vision models where the channel dimension is relatively shallower. As the channel dimension gets deeper and the spatial dimensions get narrower, Fused-IBN uses a large amount of FLOPS and parameters which substantially increases the latency cost.

\subsection{Group Convolution Based Inverted Bottlenecks}
\begin{figure}
    \centering
\includegraphics[width=3.3in]{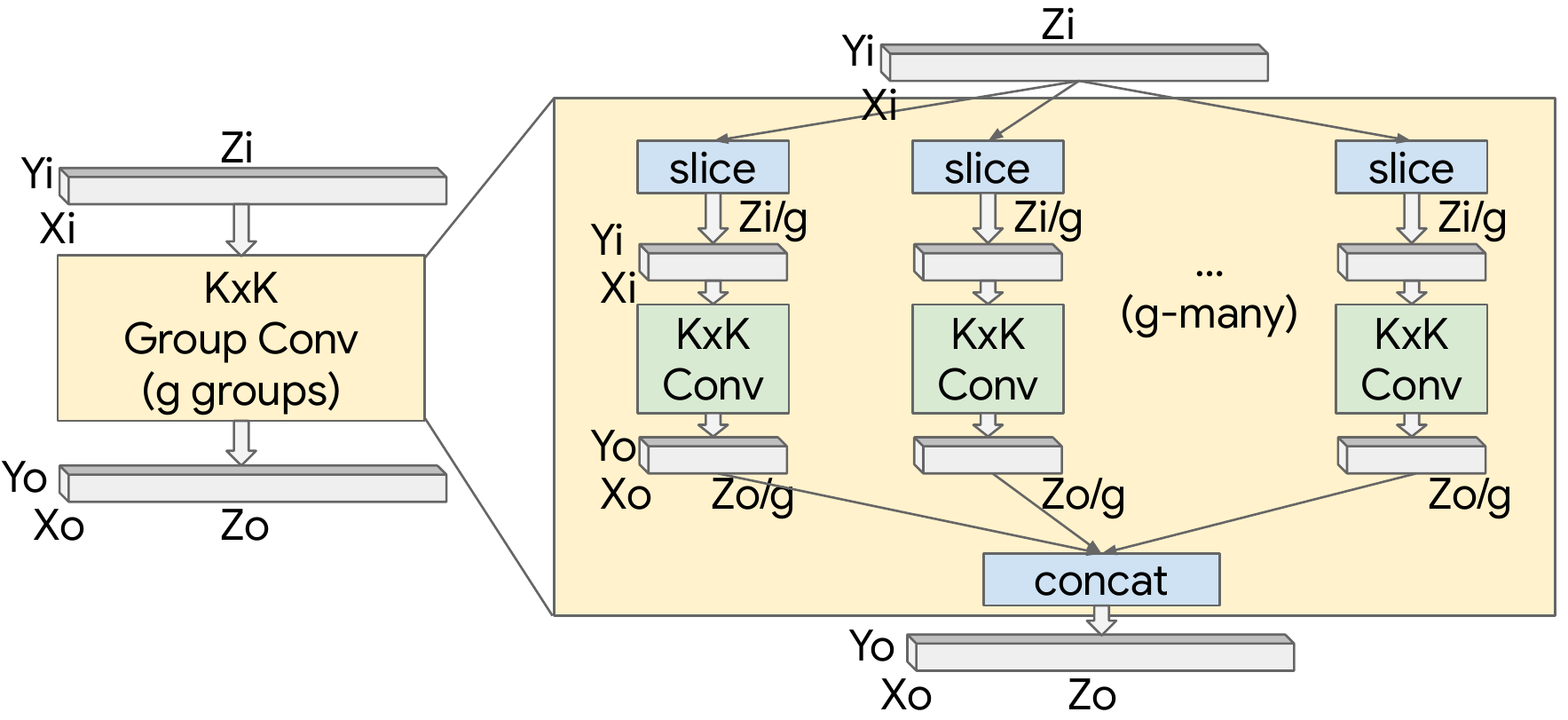}
    \caption{A $K\times K$ group convolution with $g$ groups represented as a series of regular convolutions.}
    \label{fig:gc}
\end{figure}

Group convolutions (GC) divide their input/output feature maps along the channel dimension into groups where channel-wise convolutions are limited to within each group \cite{alexnet}. A group convolution operation can be represented with a series of full convolutions applied to the groups of input and output tensors as shown in Figure~\ref{fig:gc}. GC can be considered as a generalized convolution representation such that when $g=1$ a GC becomes a regular full convolution and when $Z_i = Z_o = g$ a GC degenerates into a depthwise convolution. Therefore, one can consider the number of groups in a GC as a \emph{knob} to tune the number of parameters and operations of the convolution. This property of GC makes it a versatile tool that can be used in crafting IBN blocks. To this end, we propose GC based IBN variants to fill the gap in the neural architecture search spaces constructed solely from Depthwise and Fused IBNs. 

\begin{figure}
    \centering
\includegraphics[width=3.3in]{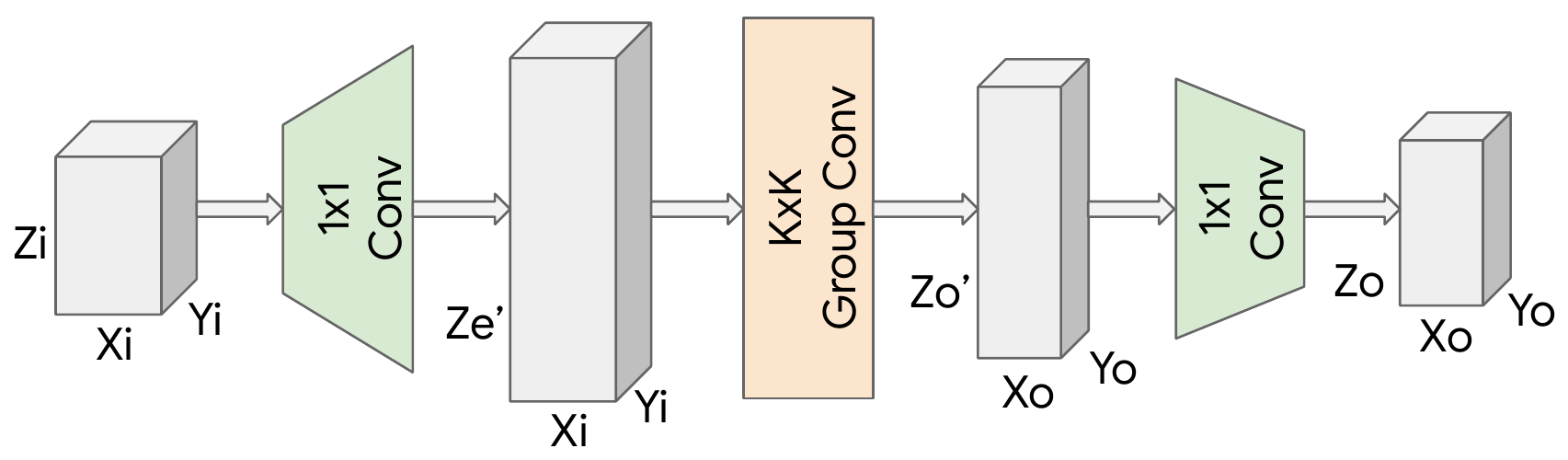}
    \caption{A generalized IBN using group convolution as the main kernel. GC can implement part of the expansion/projection since there is no constraint such that $Z_{e'}=Z_{o'}$.}
    \label{fig:ibn_gc_general}
\end{figure}

\begin{figure}
    \centering
\includegraphics[width=2.5in]{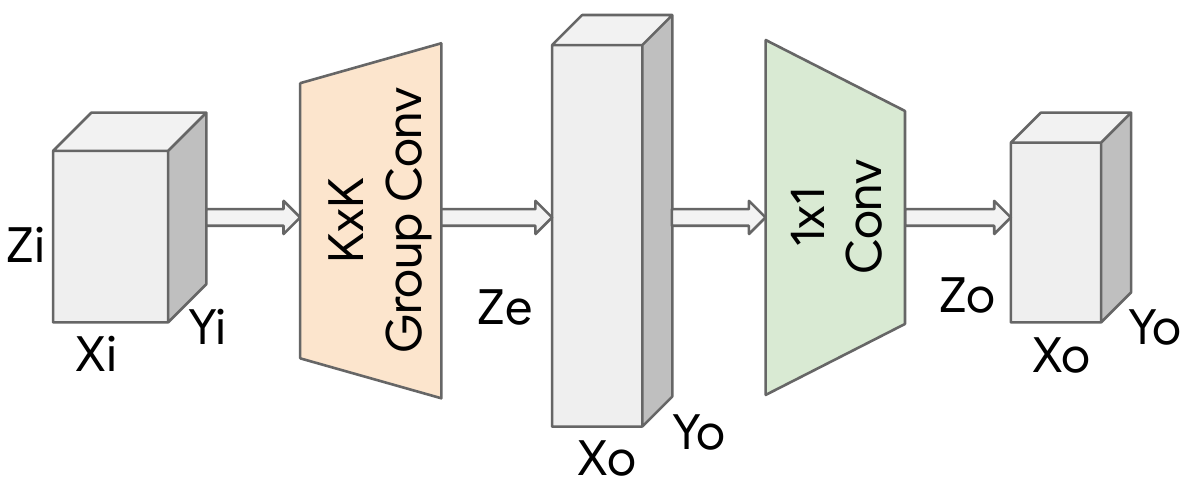}
    \caption{A special instance of Figure~\ref{fig:ibn_gc_general}: IBN using GC for the fused pointwise expansion with the main kernel (GC-IBN). (A dual block also exists where the projection is fused.)}
    \label{fig:ibn_gc}
\end{figure}

A generalized form of GC-based IBN is provided in Figure~\ref{fig:ibn_gc_general}. Firstly, GC can be used simply as a replacement of the depthwise convolution of a Depthwise-IBN to increase the total trainable parameters. However, in contrast to a depthwise convolution, GC does not constrain its input and output channel dimensions to be the same size. This allows performing a part of the channel expansion/projection using the pointwise convolutions and the remaining part by the GC kernel. For example, a total channel expansion of $m\times$, can be split into $n\times$ on pointwise convolution and $p\times$ on the GC such that $Z_{e'} = Z_i \times n$ and $Z_{o'} = Z_{e'} \times p$ where $m=n \times p$ (reverse can be applied to the projection side). Moreover, the entire expansion/projection can also be performed by the GC in which case the pointwise expansion/projection becomes ineffectual and can be eliminated (e.g. $n=1$). This instance can be considered as a Fused-IBN where the $K\times K$ convolution is replaced with a GC (Figure~\ref{fig:ibn_gc}). Due to this property we will refer to this special instance as a GC-IBN. GC-IBN provides advantages similar to the Fused-IBN as the pointwise expansion is fused into the GC kernel, yet it is more flexible thanks to the group count knob. Moreover, since the pointwise projection is kept, it provides a cross-group convolution. This allows us to avoid commonly used but hardware-unfriendly channel shuffle operations \cite{zhang2017shufflenet, wu2019fbnet}. Note that a dual of this block, where the projection is fused also exists. However, we did not include it in our search space due to its inferior performance on Edge TPUs.

\subsection{Hardware Utilization Trade-offs}
ML accelerator architectures commonly use wide single-instruction multiple-data (SIMD) execution units to extract the highest processing throughput. However, often times feeding these wide execution units from the memory system becomes the real bottleneck.

Depthwise convolutions require significantly lower number of parameters to mitigate the memory requirements. However, they fall short in utilizing the wide SIMD units of ML accelerators \cite{mobilenet_edge_paper}. An overlooked key insight related to the low utilization is the lack of the activation operand reuse in depthwise convolutions. Every input feature map element fetched from the memory is only used once when computing the output feature maps in a depthwise convolution. This puts a heavy pressure on the activation fetch bandwidth requirements, and leads to low utilization of the compute units. We make the observation that in a group convolution operation, every input feature map element fetched from the memory is reused for computing the output feature maps within its group. This is significant since this means that we can amplify the activation operand data reuse by controlling the group size as needed by the SIMD width of the hardware while requiring fewer parameters than a full convolution.

\begin{figure}
    \centering
    \includegraphics[width=3.3in]{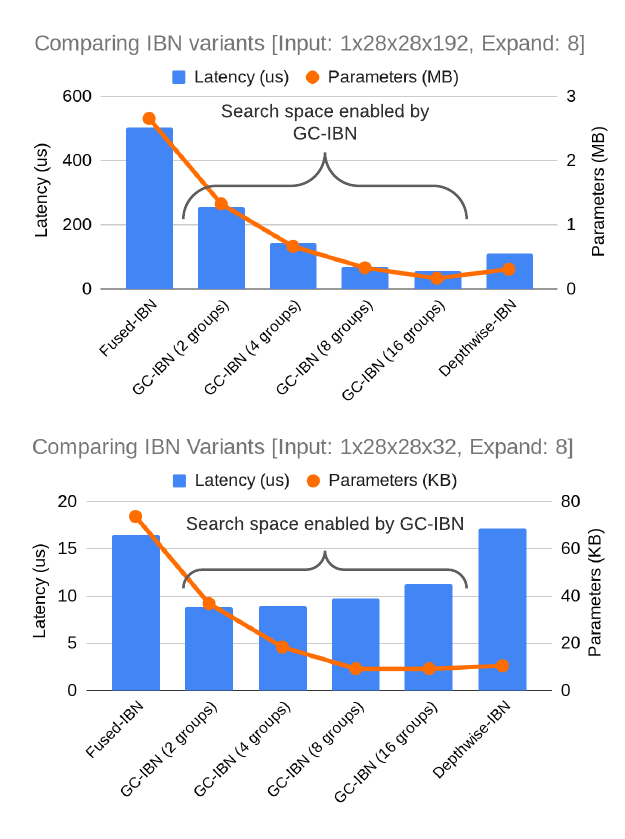}
    \caption{Executing different IBN variants for two different input sizes on Pixel 6 Tensor SoC. All IBNs use $3\times 3$ kernel size and int8 data-type.}
    \label{fig:ubench}
\end{figure}

To concretely demonstrate the computational characteristics of GC-IBNs, we leveraged the model generator as a \emph{neural block analyzer} (Section~\ref{sec:search_infra}) to generate neural nets solely based on IBN variants that can run on the Edge TPU accelerator. In Figure~\ref{fig:ubench}, first we observe that GC-IBN blocks can provide $4\times$ the trainable parameters and number of operations while having $0.5\times$ the latency cost of Depthwise-IBNs. This indeed demonstrates the importance of the data reuse in reaching high hardware utilization. We also observe that GC-IBNs hardware utilization can be closer to the Fused-IBN blocks especially with smaller number of groups (hence larger group sizes). With smaller group sizes we start to lose data reuse and hit diminishing returns. Finally, we observe that the latency vs. trainable parameter count trade-offs are highly dependent on the tensor shapes and choosing the optimal IBN variant and its configuration (e.g., group size) is not a straight-forward task which calls for an automated exploration methodology using a neural architecture search (NAS).

With the inclusion of the proposed IBN variants the neural architecture search space becomes extremely large.
Although choosing the optimal blocks that will maximize the model quality with a given latency target is a very difficult task to perform by hand, we observe that some block choices can be inherently sub-optimal for certain places in the neural network topology. For example, in the later stages of the neural network with the growth of the channel dimension and the reduction of the spatial dimension, parameter reuse drops significantly and Fused-IBNs become much less efficient. We leverage the neural block analyzer to alleviate the search space size and filter out such choices by carefully analyzing the IBN's performance characteristics.

%% file: 4_models.tex
\section{Edge TPU Optimized Models}
\label{sec:models}
In this section, we use the proposed infrastructure on the search spaces including the proposed GC-based IBNs and target the Edge TPU ML accelerator in the Google Tensor mobile SoC \cite{tensor_soc} for the on-device ML tasks identified by MLPerf Mobile Inference suite \cite{mlperf_mobile} which includes image classification, object detection, semantic segmentation and natural language processing.

\input{4_1_classification}

\input{4_2_segmentation}

\input{4_3_detection}

\input{4_4_nlp}

\input{4_5_energy}

%% file: 4_1_classification.tex
\subsection{Image Classification}
\label{sec:classify}
We start from a model topology similar to MobileNet/MobileDets due to their efficiency for mobile platforms including TPUs which is our primary target \cite{mobiledet_cvpr_xiong, mobilenet_edge_paper, mobilenet_edge_blog}. In the search space, we include the IBN variants described in Section \ref{sec:search_space} including Depthwise, Fused and GC-IBNs. We include residual skip connections over the IBN blocks with unit stride but omit them for the blocks that use stride $>1$. We also omit swish non-linearity and the squeeze-and-excite blocks which are known to be less efficient on edge ML accelerators.
As mentioned previously, we fine-tune the search space by filtering the IBN variants with consistently sub-optimal performance characteristics at certain blocks of the model topology instead of including all variants globally. Furthermore, considering the wide SIMD engines of ML accelerators, we pick a minimum group size of 32 and omit GC-IBNs with smaller group sizes (i.e. larger group counts) based on the neural block analyzer (Section~\ref{sec:search_space}).

Using this search space, we target the Edge TPU ML accelerator in the Google Pixel 6 Tensor SoC. We search for 5 different models with progressively increasing latency budgets which are named as Tiny, XS, S, M, L variants of MobileNetEdgeTPUv2. Accuracy vs. latency trade-offs provided by these models after post-training quantization to int8 datatype are provided in Figure~\ref{fig:classify_tpu} in comparison to other state-of-the-art (SOTA) mobile models. We observe that MobileNetEdgeTPUv2 model family outperforms even the prior SOTA MobileNetEdgeTPU models that are optimized for the Edge TPUs.

\begin{figure}
    \centering
    \includegraphics[width=3.33in]{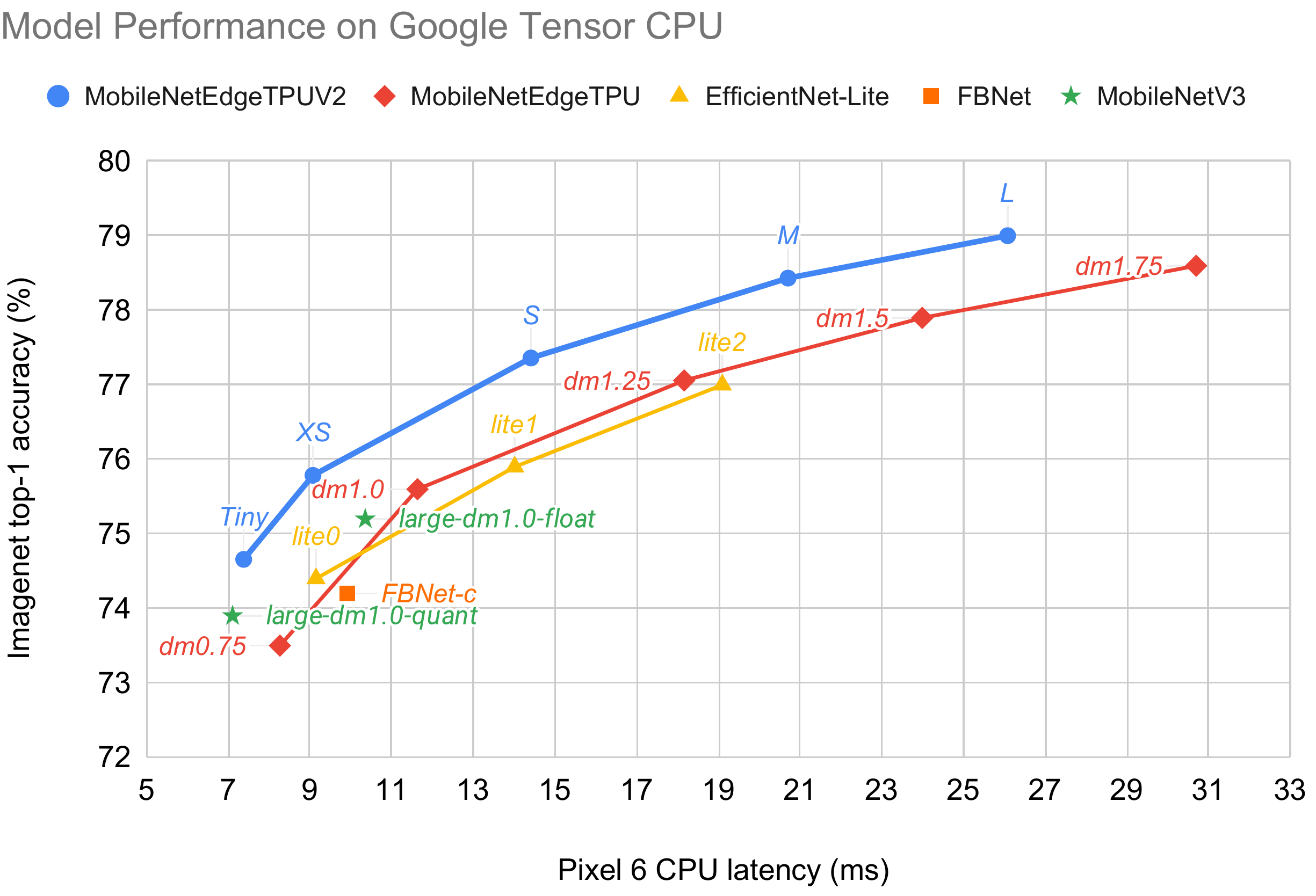}
    \caption{MobileNetEdgeTPUV2 models also demonstrate better accuracy-latency trade-off on Google Tensor CPU.}
    \label{fig:classify_cpu}
\end{figure}

Our primary optimization target is the TPU accelerator, however our search space includes operations that also run well on mobile CPUs. Moreover, we implement GC using functionally equivalent series of commonly used ML primitives (slice, full convolution, concatenation) as shown in Figure~\ref{fig:gc}, so that various platform compilers can efficiently support them since the native GC support may be missing. Also GC-IBNs tend to have fewer operations than Fused-IBNs and mobile CPUs show a stronger correlation between the number of operations in the neural network and latency. As a result, in Figure~\ref{fig:classify_cpu} we observe that when executed on the Google Tensor CPU, MobilenetEdgeTPUV2 family also outperform other SOTA models.

%% file: 4_2_segmentation.tex
\subsection{Semantic Segmentation}
\label{sec:segment}
Many vision models consist of two components, the base feature extractor for understanding general features of the image, and the head for understanding domain-specific features, such as semantic segmentation. For feature extraction, we start from a model topology similar to EfficientNet~\cite{EfficientNet2019} with IBN variants described in Section~\ref{sec:search_space}. We use a MobileNetEdgeTPUv2 classification model coupled with the DeepLabv3~\cite{chen2017rethinking} segmentation head as our baseline model and find that it improves the quality of on-device segmentation.

\begin{figure}
    \centering
    \includegraphics[width=3.33in]{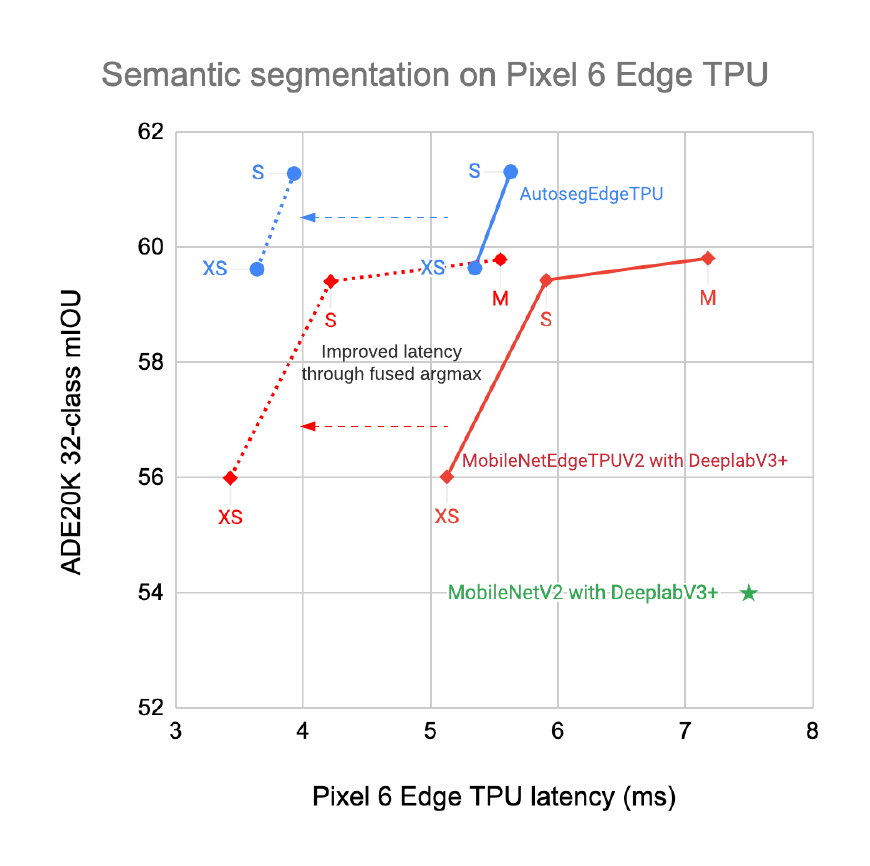}
    \caption{Segmentation model performance on Edge TPU.}
    \label{fig:seg}
\end{figure}

To further improve the segmentation model quality, we use the bidirectional feature pyramid network (BiFPN)~\cite{tan2020efficientdet} as the segmentation head, which performs weighted fusion of different features extracted by the feature extractor. Using NAS we find the optimal configuration of blocks in both the feature extractor and the BiFPN head. 
Specifically, we search for the kernel size from \{3, 5, 7\} for each IBN layer, and we also search for the expansion ratio from \{3, 6\} for each block except for the first one, which has the default expansion ratio of 1. In addition, we apply a channel multiplier that is among \{1/2, 1/4, 1, 3/4, 2\} to scale the model up and down. In the BiFPN head, we search over the number of repeats and minimum feature level, which produces trade-offs between accuracy and latency. 
The resulting models, named Autoseg-EdgeTPU, produce even higher-quality segmentation results, while also running faster (Figure~\ref{fig:seg}).

The final layers of the segmentation model contribute significantly to the overall latency, mainly due to the operations involved in generating a high resolution segmentation map. To optimize the latency on TPU, we introduce an approximate method for generating the high resolution segmentation map that reduces the memory requirement and provides a nearly 1.5x speedup, without significantly impacting the segmentation quality. 

%% file: 4_3_detection.tex
\subsection{Object Detection}
\label{sec:detect}
Modern one stage object detection architectures typically produce one or more feature maps from the input image and use either an anchorless detection head such as the CenterNet\cite{centernet_cvpr_zhou} or anchor based detection head such as the SSD\cite{ssd_eccv_liu}. In the past, the classic process to design architectures for either of these two types of object detectors requires choosing a backbone network such as the MobileNets\cite{mobilenet_cvpr_howard, mobilenet2_cvpr_sandler} for low latency applications or ResNets\cite{resnet_cvpr_he} for accurate applications.

There are various methods to fuse together the feature maps from different endpoints of the backbone, such as FPN\cite{fpn_cvpr_lin} which iteratively includes more low level information into the feature map as it upsamples the top feature map.

We notice that most of the classic object detection architectures allocate more than $70\%$ of the total budget to the backbone area of the network while limiting the feature map fusion to less than $30\%$. We want to explore if rebalancing such allocation could lead to a better detection architecture. Also, recent NAS works such as MnasFPN\cite{mnas_fpn_cvpr_chen} introduces a non-trivial connection pattern to fuse feature maps from different endpoints in the backbone network. We want to utilize the success in such connection patterns as we design our object detection architecture.

With this in mind, we have created the Spaghetti Search Space, aptly named for the spaghetti-like connections between architecture blocks. For the COCO Object Detection Task\cite{coco_lin}, the search space consists of a stem node and 12 main blocks, each with the choice of between 2-4 layers. 6 blocks form the backbone whilst the other 6 form the head. Blocks in the head use the MnasFPN\cite{mnas_fpn_cvpr_chen} connection pattern. Each layer may consist of depthwise separable cnvolutions\cite{depthwise_sep_conv_chollet}, Inverted Bottleneck blocks (IBN)\cite{mobilenet2_cvpr_sandler} and Grouped Convolution based IBNs (GC-IBN).

\begin{figure}[h!]
\centering
\includegraphics[scale=0.34]{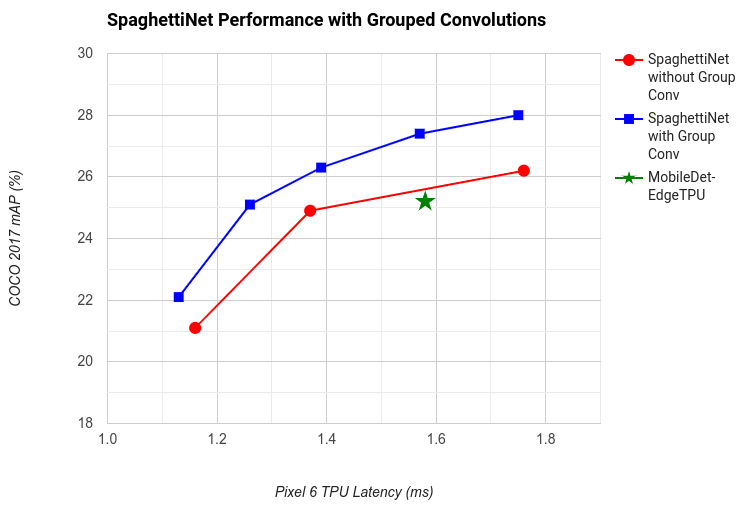}
\caption{The performance of SpaghettNet models compared to MobileDet-EdgeTPU. When GC-IBN blocks are incorporated into the Spaghetti search space, it achieves +2.2\% mAP more than MobileDet-EdgeTPU at the same latency.}
\label{fig:spaghettinet_performance}
\end{figure}

As seen in Figure ~\ref{fig:spaghettinet_performance}, models found with this search space outperform MobileDet-EdgeTPU\cite{mobiledet_cvpr_xiong}, current state-of-the-art detection models targeting Edge TPU platform. To verify the usefulness of GC-IBNs we remove GC-IBNs from the search space, and observe that the optimal models perform similarly to MobileDet-EdgeTPU, 
which demonstrates that the proposed search space provides more efficient options compared to the existing 
depthwise and full convolution based IBNs.

%% file: 4_4_nlp.tex
\subsection{Natural Language Processing}
\label{sec:nlp}
Deploying low-latency, high-quality transformer based language models on-device is highly desirable, and can potentially benefit multiple applications such as automatic speech recognition (ASR), translation, sentence auto-completion, and even some vision tasks \cite{dosovitskiy2020image}. 
While we mainly focused on vision tasks so far, the NAS infrastructure is domain agnostic, and can be easily extended to applications beyond vision, such as BERT \cite{devlin2018bert} variant of language models.

\begin{figure}[h!]
\centering
\includegraphics[width=3.33in]{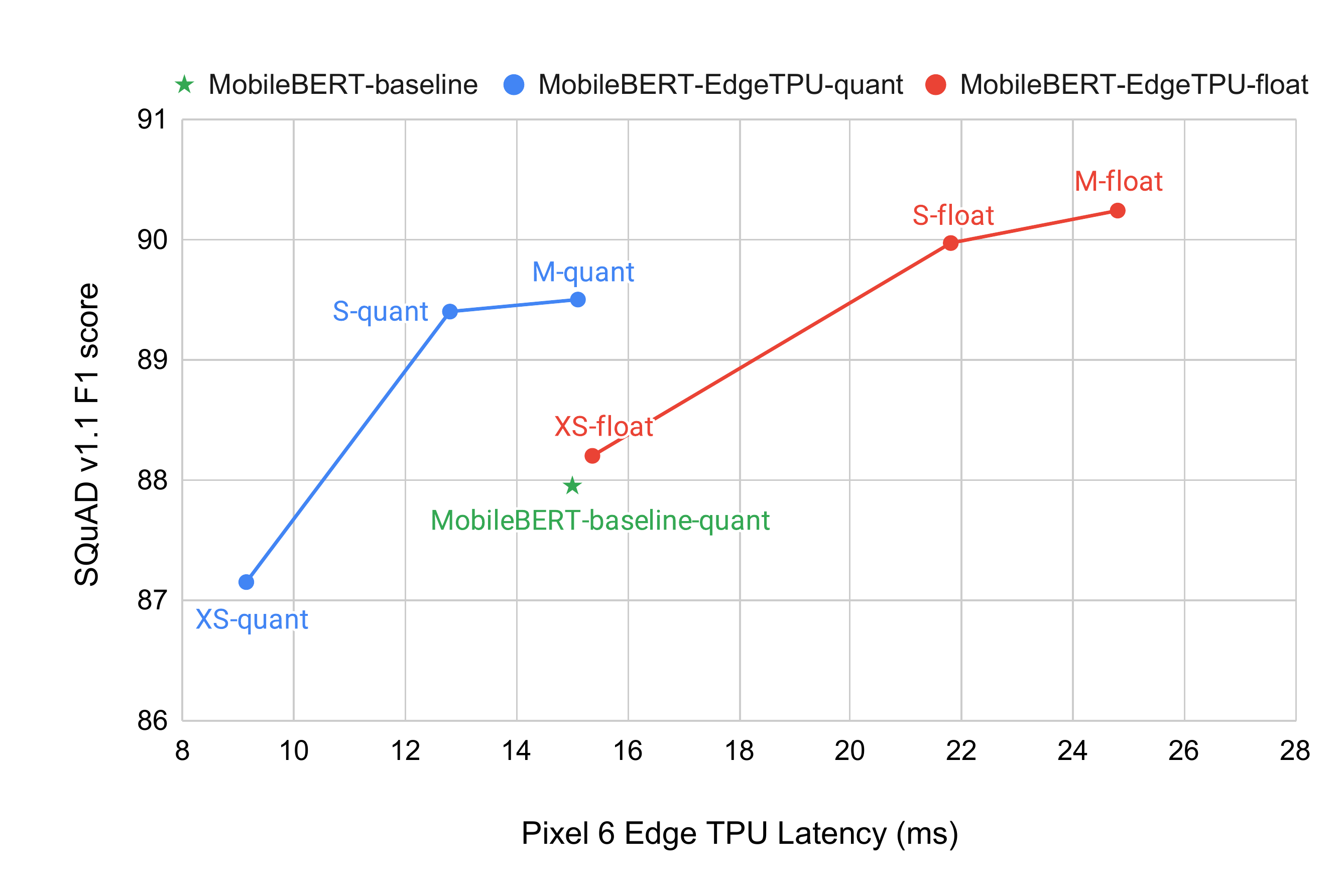}
\caption{Performance of MobileBERT-EdgeTPU models on the SQuAD v1.1 dataset.}
\label{fig:mobilebert-edgetpu-performance}
\end{figure}

Due to the limitations on weight sharing support in TuNAS for transformers,
we simply use a multi-trial approach exploiting the flexibility of the proposed NAS infrastructure.
Named as Mobilebert-EdgeTPU, we set up our NLP model architecture search space based on MobileBERT \cite{sun2020mobilebert} and leverage the proposed NAS framework to find models with up to 2x better Edge TPU hardware utilization. 
With higher utilization, we are able to bring larger and more accurate models on chip, and meanwhile the models can still outperform the baseline MobileBERT latency.
To complement the model generator for multi-trial NAS,
we developed a customized knowledge distillation based training pipeline
to quickly assess the generated model's quality without full training during search. 
The final model is fully trained. 
As shown in figure \ref{fig:mobilebert-edgetpu-performance}, the quantized MobileBERT-EdgeTPU models establish a new pareto-frontier for the question answering tasks and also exceed the accuracy of the float $BERT_{base}$ \cite{devlin2018bert} model, a 400MB+ model in float32 precision which is too large to run on edge devices. 

As an alternative to the quant models, we also provide a set of Edge TPU friendly float models, as shown in figure \ref{fig:mobilebert-edgetpu-performance}. Notably, the float MobileBERT-EdgeTPU-M model yields accuracy that is even comparable to the $BERT_{large}$\cite{devlin2018bert}, which has 1.3GB model size in float32 precision. Quantization now becomes an optional optimization rather than a prerequisite, which can greatly benefit use cases where quantization is infeasible or introduce large accuracy deterioration, and potentially reduce the time-to-market.

%% file: 4_5_energy.tex
\subsection{Energy Efficiency}
\label{sec:energy}
\begin{figure}
    \centering
    \includegraphics[width=3.33in]{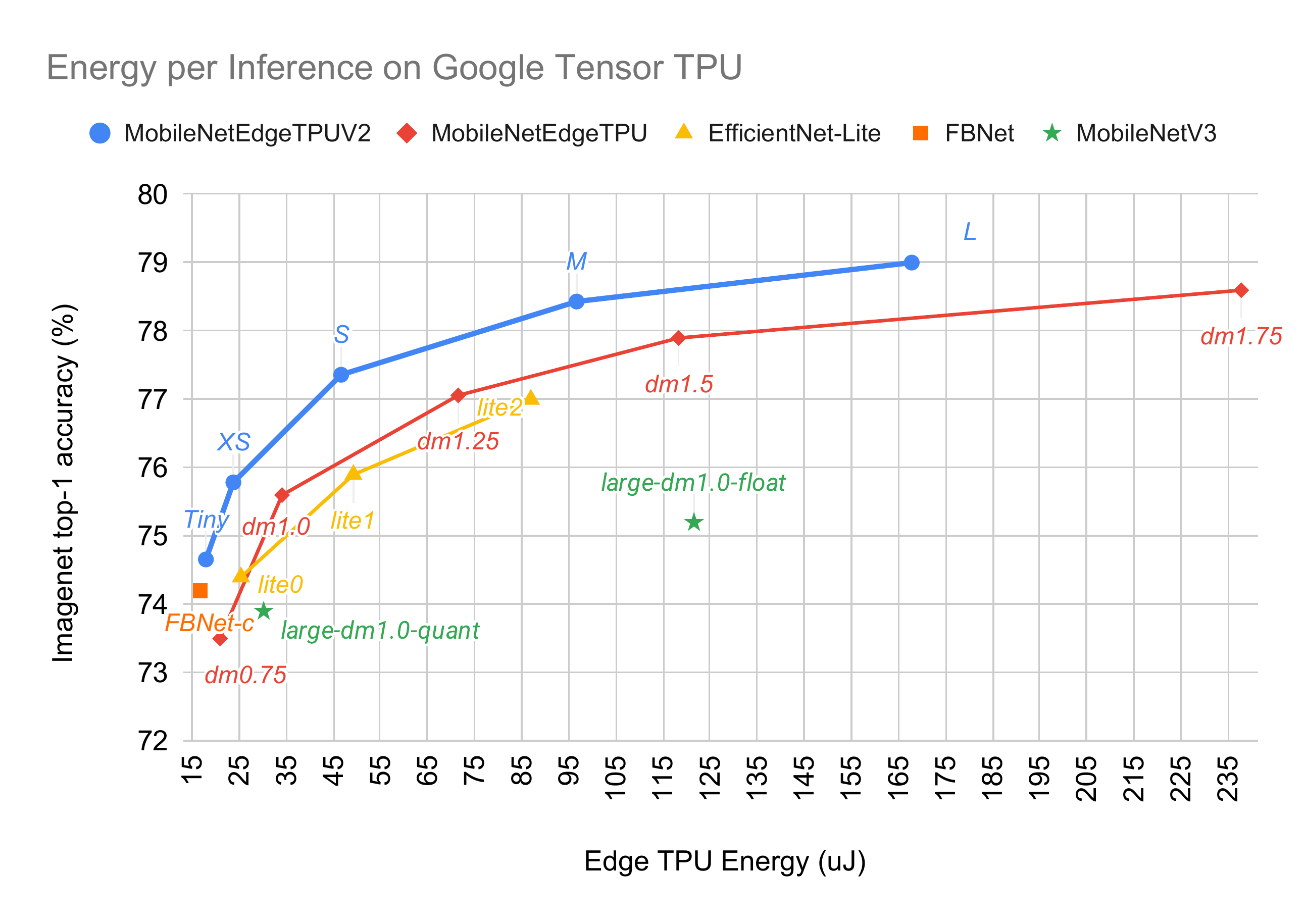}
    \caption{On-device energy per inference measurements when running the models on Edge TPU.}
    \label{fig:energy}
\end{figure}

As the energy consumption is critical for on-device ML use cases,
we also setup an energy measurement harness and benchmark our models.
Our benchmarking setup uses the nominal device settings and 
runs the models at 30 inferences per second.
We first measure the average power while the TPU is idle. 
Then, we subtract the idle power from the average power measured 
when the TPU is running the model for 100 inferences to find the active TPU power consumption.
Finally, this power consumption rate is multiplied by the model latency to find 
the energy consumed per inference.
Figure~\ref{fig:energy} demonstrates that the energy efficiency trends are similar 
to the latency measurements. 
This is expected since the efficient utilization of the hardware not only improves performance 
but also minimizes the use of inefficient operations and reduces the energy consumption.
For the other targeted tasks we observe similar trends but for brevity we only report the 
image classification results.

%% file: 6_conclusion.tex
\section{Conclusion}
\label{sec:conc}
In this work we target optimizing various on-device ML tasks on edge ML accelerators.
We propose flexible inverted-bottleneck (IBN) variants using group convolutions (GC) and design 
search spaces including these blocks.
GC based IBNs opens up the search space between depthwise and full convolution based IBNs,
and create unique opportunities for efficient execution on ML accelerators.
To easily find optimized models for various on-device ML tasks, we propose a scalable NAS-enabling infrastructure that decouples cost evaluation, neural search space design. 
Using this infrastructure with the proposed search spaces and targeting a state-of-the-art mobile SoC platform Google Tensor TPU, we demonstrate significant improvements in quality, latency and energy metrics for mobile ML tasks including computer vision (classification, detection, segmentation) and natural language processing (NLP).